\documentclass[12pt]{article}

\usepackage{hyperref}
\usepackage{amssymb}
\usepackage{graphicx}

\begin{document}

\renewcommand{\abstractname}{\hfill}

\newpage
\pagenumbering{arabic}

{\Large \bf Particles of negative and zero energy

\vspace{4pt}
in black holes and cosmological models}

\vspace{11pt}
{\bf
Andrey A. Grib${}^{1,2}$ and Yuri V. Pavlov${}^{3,*}$
}
\begin{abstract}
${}^{1}$ Theoretical Physics and Astronomy Department, The Herzen  University,
48~Moika, St.\,Petersburg, 191186, Russia;  andrei\_grib@mail.ru

${}^{2}$ A. Friedmann Laboratory for Theoretical Physics, St.\,Petersburg, Russia

${}^{3}$ Institute of Problems in Mechanical Engineering of
Russian Academy of Sciences,
61 Bolshoy, V.O., St.\,Petersburg, 199178, Russia


${}^{*}$ Correspondence: yuri.pavlov@mail.ru
\end{abstract}

    \begin{abstract}
\noindent
{\bf Abstract:}
    Particles with negative energies are considered for three different cases:
inside of horizon of Schwarzschild black hole,
Milne's coordinates in flat Minkowski space-time (Milne's universe using
nonsynchronous coordinates)
and in cosmological G\"{o}del model of the rotating universe.
    It is shown that differently from the G\"{o}del model with
nondiagonal term where it occurs that negative energies are impossible
they are present in all other cases considered in the paper.
    Particles with zero energy are also possible in first two cases.

\vspace{7pt}
\noindent
{\bf Key words:} \ Negative energy; Zero energy; Black hole;
Milne universe; G\"{o}del universe
\end{abstract}

\section{\normalsize Introduction}
\label{secIn}

\hspace{\parindent}

    In 1922 Alexander Friedmann in Petrograd, Russia predicted an expansion
of the Universe.
    Today Friedmann's model of the Universe is called the Standard Model.
    Many observations confirm this model to be correct.
    The new world of galaxies and the stages of the Universe evolution were
discovered.
    New phenomena, such as black holes, relict radiation and many others,
are actively investigated.
    Trajectories of particles with negative and zero energy are examples
of such new phenomena.
    The possibility of relativistic particles with negative energy is
important because it makes possible to get large energy in interactions or
decays of bodies.
    The simple example of this situation was proposed by R.\,Penrose in
the case of the decays of particles in ergosphere of rotating black
holes~\cite{Penrose69,PenroseFloyd71}.

    It looks that in order to have negative energy of the relativistic
particle with nonzero mass one must have very strong external field leading
to large potential energy as it is in the case for rotating black holes.
    However it is well known that the value of the energy depends on the
choice of the reference frame and the time coordinate or Killing vector in
case of conserved energy.
    It leads to the situation that states with negative energies in
relativistic case occur in case of rotating coordinate system out of the static
limit~\cite{GribPavlov2016d},
where the effect analogical to Penrose effect turns out to be
observable~\cite{GribPavlov2019},
and in nonsynchronous coordinate system in
cosmology~\cite{GribPavlov2019h,GribPavlov2020}.

    It seems from these examples that negative energies arise in case of
the existence of nondiagonal terms in metrical tensor (gravymagnetism)
but in this paper we show that in G\"{o}del universe in spite of the
presence of such terms negative energies are absent.
    The negative energies are present in cases of the movement inside of
the horizon of the Schwarzschild black hole and in Milne's universe where
nondiagonal terms are present in nonsynchronous coordinate frame.

\vspace{4mm}
\section{\normalsize Negative energy in nonrotatitng black hole}
\label{secSch}

\hspace{\parindent}
    Nonrotating black hole of mass~$M$ in Schwarzschild coordinates
is described by metric
    \begin{equation}
d s^2 = \left( 1 \!-\! \frac{r_g}{r} \right) c^2 d t^2 - \frac{d r^2}{1 \!-\!
\frac{r_g}{r} } - r^2 \! \left( d \theta^2 \!+ \sin^2\! \theta \, d\varphi^2 \right),
\label{gSch1}
\end{equation}
    where $ r_g = 2 G M /c^2 $ is the gravitational radius of the black hole,
$G $ is gravitational constant, $c$~is the light velocity.
    Geodesic complete space-time of the nonrotating black hole can be
described in Kruskal--Szekeres coordinates,
$ \{ u, v \} \in (-\infty, +\infty) $, which in region $u > |v| \ge 0 $
are connected with Schwarzschild coordinate in $r > r_g$
in the following way
    \begin{equation}
u = \sqrt{ \frac{r}{r_g} - 1 } \exp \left( \frac{r}{2 r_g} \right)
\cosh \frac{ct}{2 r_g}, \ \ \
v = \sqrt{ \frac{r}{r_g} - 1 } \exp \left( \frac{r}{2 r_g} \right)
\sinh \frac{ct}{2 r_g}.
\label{gSch2}
\end{equation}
    For $r < r_g$ and $ v > |u| \ge 0 $
the transformation from Schwarzschild coordinate in the
Kruskal--Szekeres coordinates has the form
    \begin{equation}
u = \sqrt{ 1 - \frac{r}{r_g} } \exp \left( \frac{r}{2 r_g} \right)
\sinh \frac{ct}{2 r_g}, \ \ \
v = \sqrt{ 1 - \frac{r}{r_g} } \exp \left( \frac{r}{2 r_g} \right)
\cosh \frac{ct}{2 r_g}.
\label{gSch2d}
\end{equation}
    Schwarzschild coordinates are singular at $r=r_g$.
    Their connection with Kruskal--Szekeres coordinates for other $u,v$,
see in~\cite{MTW}, Sec.~31.5.

    Any possible movement of physical bodies and particles must
satisfy the condition $ {d s^2 >0} $ leading to
    \begin{equation}
r > r_g \ \ \Rightarrow \ \
\left| \frac{d r}{d t} \right| \le c \left( 1 - \frac{r_g}{r} \right),
\label{gSch3}
\end{equation}
    \begin{equation}
r < r_g \ \ \Rightarrow \ \
\left| \frac{d r}{d t} \right| \ge c \left( \frac{r_g}{r} - 1 \right).
\label{gSch4}
\end{equation}
    For $r<r_g$ the coordinate $ct$ is space and $r/c$ --- time coordinate.

    Geodesic equations in Schwarzschild coordinates in the plane $\theta = 0$
are~\cite{Chandrasekhar}
    \begin{equation}
\frac{d t}{d \lambda} = \frac{r}{r-r_g} \frac{E}{c^2}, \ \ \
\frac{d \varphi}{d \lambda} = \frac{J}{r^2}, \ \ \
\left( \frac{d r}{d \lambda} \right)^2 = \frac{E^2}{c^2} +
\frac{r_g-r}{r^3} J^2 + \frac{r_g - r}{r} m^2 c^2,
\label{gSch5}
\end{equation}
     where $E$ is the energy of a moving particle
(measured by the static observer in $r$, $\theta$, $\phi$ coordinates),
$J$ is the conserved projection of the particle angular momentum
on the axis orthogonal to the plane of motion, $m$ is the particle mass,
$\lambda$ is affine parameter on geodesic.
    For massive particle $\lambda= \tau/m$, where $\tau$ is
the proper time.

    In external region of the black hole ($r>r_g$) for any physical object
the time coordinate~$t$ is always increasing and so the energy~$E$ of
the particle is positive (see~(\ref{gSch5}).
    Inside the horizon of the black ($r<r_g$), where~$t$ is space like
($g_{tt} <0$) one has movement as in increasing as in decreasing~$t$.
    As it is seen from the first formula in~(\ref{gSch5}) for a particle
moving inside the horizon in the direction of decreasing of the
coordinate~$t$ the energy~$E$ of the particle will be positive
while for increasing coordinate~$t$ the energy $E$ is negative.
    For constant~$t$ inside the black hole $ E=0 $ due to
formula~(\ref{gSch5}).

    Surely~$t$ inside of the black hole is space like and $E$ is proportional
to the $t$-component of the momentum.
    Inside black hole one can use other reference frames and
corresponding energies~\cite{GribPavlov2010NE}.
    But for the observer outside of the black hole the conserved~$E$ along
all trajectories of the free fall is equal
(see formula (88.9) in~\cite{LL_II}) to
    \begin{equation}
E= m c^2 \sqrt{ \left( 1 - \frac{r_g}{r} \right) \left/
\left( 1 - \frac{ \mathbf{v}^2 }{c^2} \right) \right.},
\label{gSch6}
\end{equation}
    where $\mathbf{v}$ is the velocity measured by the observer at rest
in the Schwarzschild coordinates.
    So we can call~$E$ inside the black hole following~\cite{MTW},
as ``energy at infinity''.
    The discussion of other ways to determine energy within the horizon and
the movement of particles there, see, for example, in
the articles~\cite{GribPavlov2010NE}, \cite{Toporen}--\cite{Radosz}.

\begin{figure*}[ht]
\centering
\includegraphics[width=70mm]{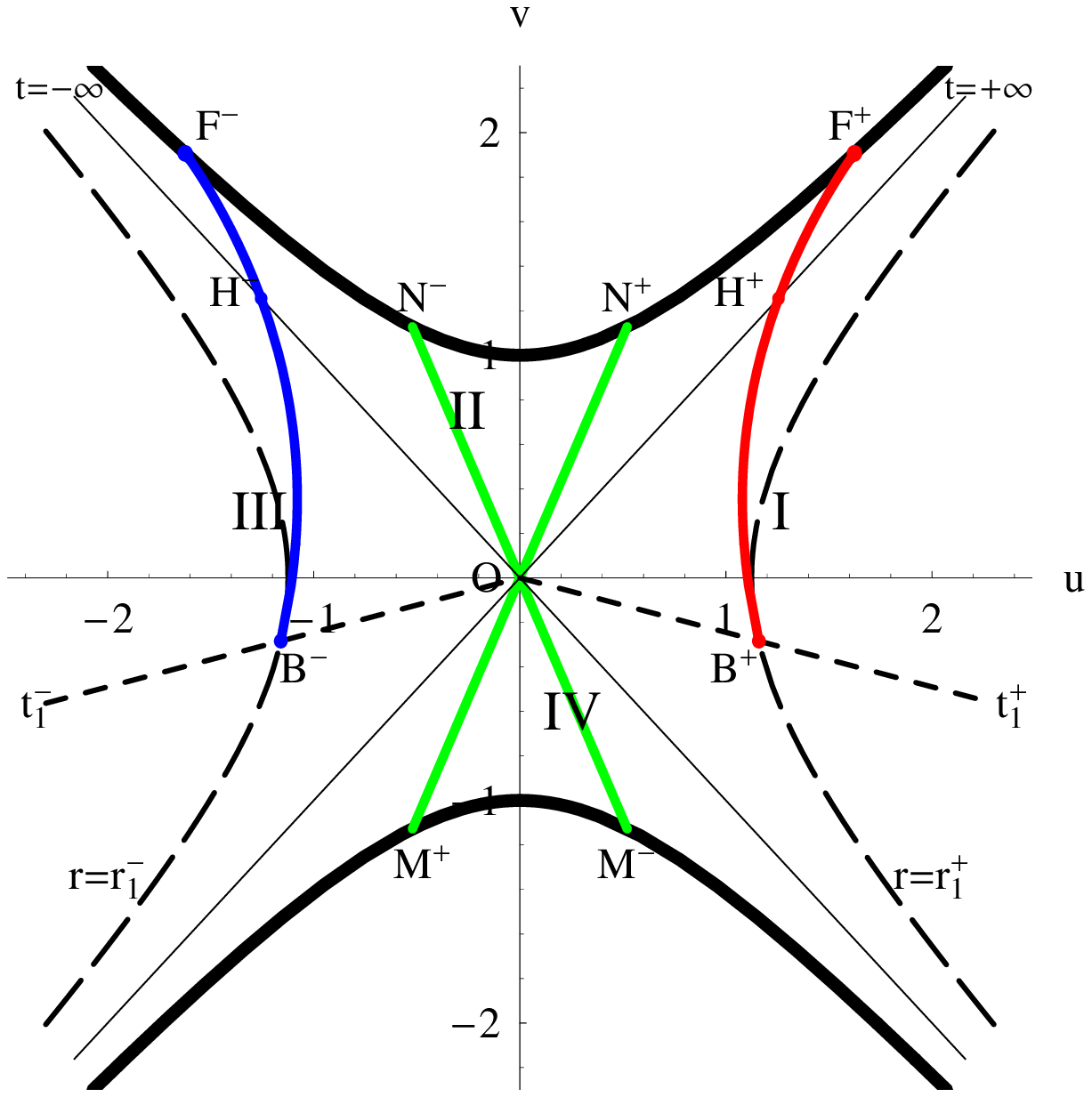} \ \ \ \ \ \ \
\includegraphics[width=70mm]{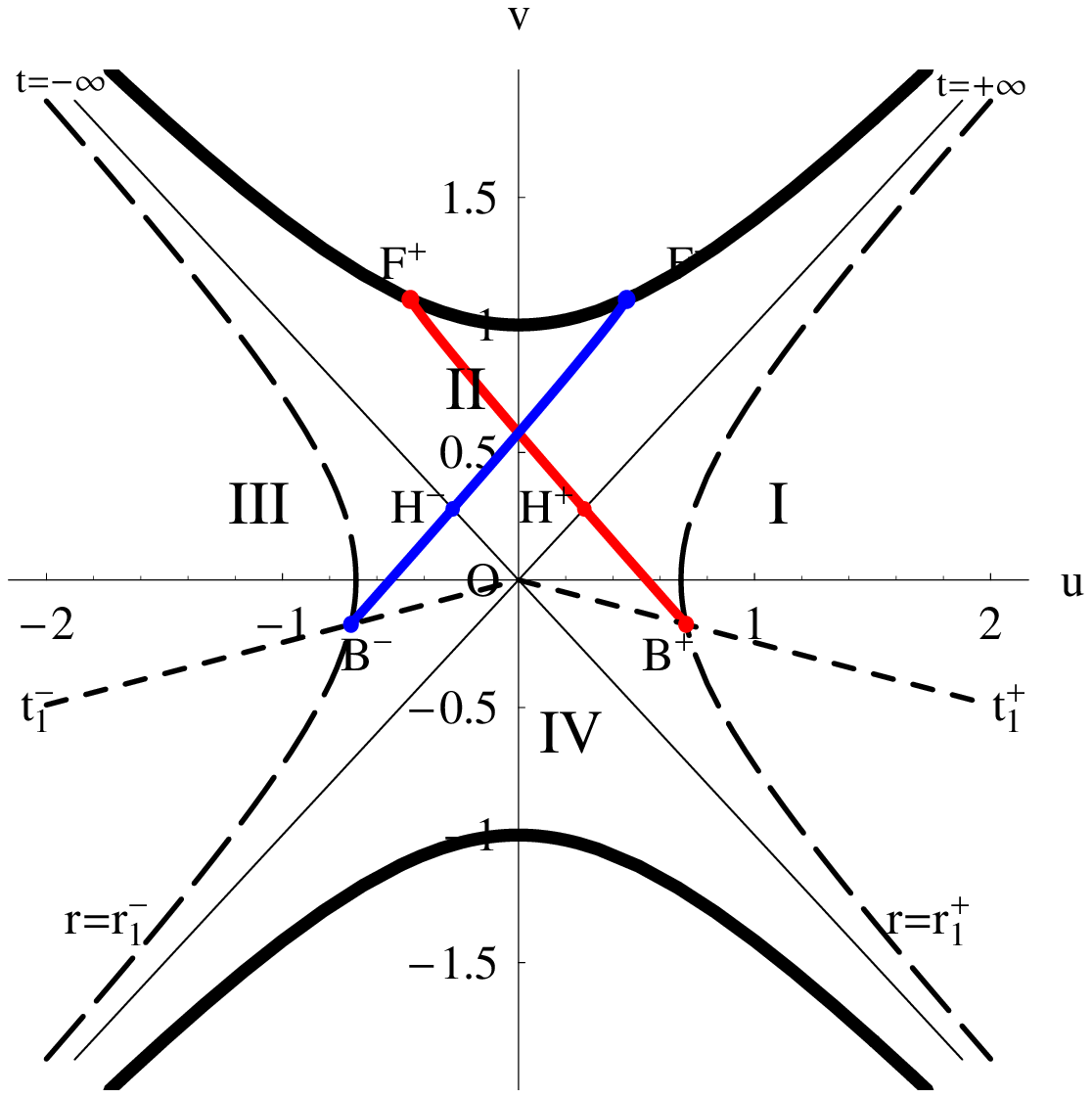}
\caption{Trajectories of particles with positive ($B^+H^+F^+$),
zero ($M^\pm ON^\pm$) and negative ($B^-H^-F^-$) energy,
$t_1^\pm = \mp 0.5 r_g/c$:
On the left one can see falling from the rest at $ |E|= 0.5 m c^2$,
on the right with $ |E|= m c^2$ with the corresponding initial velocity
from the point $r=1.15 r_g$.
On lines ($M^\pm ON^\pm$) the coordinate $t= \pm r_g/c$.}
\label{FigSch}
\end{figure*}

    On Fig.~\ref{FigSch} the trajectories for radial movement with positive,
zero and negative energies in Kruskal--Szekeres coordinates are represented
by red, green and blue lines.
    As one can see from~(\ref{gSch2}), (\ref{gSch2d}) the coordinate lines
of constant $t$ in Kruskal--Szekeres coordinates are straight lines through
the origin of coordinates.
    In region II coordinates $t$ decreases for moving from $H^+$ to $F^+$
(positive $E$) and increases for moving from $H^-$ to $F^-$
(negative value of~$E$).
    Direct lines ($M^\pm ON^\pm$) correspond to constant $t=\pm r_g/c$ and
therefore $E=0$.

    Let us consider the problem of back influence of falling particles
on metric of the black hole space-time.
    For macroscopic bodies with 4-velocity $(u^i)$, with
the energy density~$\varepsilon$ and pressure~$p$ in space-time with
metric~$g_{ik}$ the energy-momentum density tensor is~\cite{LL_II}
    \begin{equation}
T_{ik} = (\varepsilon + p ) u_i u_k - p g_{ik},
\label{gSch7}
\end{equation}
    $i,k=0,1,2,3$.
    The trace of the energy-momentum tensor
    \begin{equation}
T_{i}^i = \varepsilon - 3 p
\label{gSch8d}
\end{equation}
     is invariant and it will be negative for $ \varepsilon - 3 p <0$,
in particular, for dust like matter ($p=0$) with negative energy
$\varepsilon < 0$.
    The back influence of falling particles with negative energy
will be determined by the such energy-momentum tensor in the right hand side
of Einstein equations.
    Notion of the existence of particles with negative energies
as it is known was used by S.\,Hawking to predict Hawking effect
for black holes~\cite{Hawking75}.

    The discussion of other ways to determine energy within the horizon and the movement of particles there, see, for example, in works

\vspace{4mm}
\section{\normalsize Negative and zero energies in flat space-time}
\label{secFC}

\hspace{\parindent}
    The geodesic lines equations can be obtained for space-time with
metric $g_{ik}$ from the Lagrangian
    \begin{equation}
L= \frac{g_{ik}}{2} \frac{d x^i}{d \lambda} \frac{d x^k}{d \lambda},
\label{Le1}
\end{equation}
    where $ \lambda $ is the affine parameter on the
geodesic~\cite{Chandrasekhar}.
    The energy of the particle~$E$ is equal to the zero covariant component
of the momentum  $(p_i)$ multiplied on the light velocity
    \begin{equation}
p_i = \frac{\partial L}{\partial \left( \frac{d x^i}{d \lambda}\right) } =
g_{ik} \frac{d x^k}{d \lambda},
\label{Le2}
\end{equation}
    \begin{equation}
E = c p_0 = c g_{0k} \frac{d x^k}{d \lambda}.
\label{Le3}
\end{equation}
    Defining the affine parameter for the massive particle
as $ \lambda = \tau / m $, where $ \tau $ is the proper time of the moving
particle one obtains
    \begin{equation}
p_i p^i = m^2 c^2
\label{Lee3}
\end{equation}
    and the energy of the particle is
    \begin{equation}
E = m c g_{0k} \frac{d x^k}{d \tau}.
\label{Le4}
\end{equation}

    Using notation $(\zeta^i) = (1,0,0,0)$ for the translation in time
coordinate generator one can write~(\ref{Le3}) for the energy of
the particle as
    \begin{equation}
E = c (p,\zeta).
\label{Le5}
\end{equation}
    If the metric components don't depend on the time coordinate~$x^0$,
then $\zeta$ is the time like Killing vector and the energy $E$
is conserved on the geodesic.
    For time like vector~$\zeta$ and massive particle one has~\cite{LPPT}
    \begin{equation}
\sqrt{(\zeta,\zeta)} \le \frac{E}{ m c^2 } < + \infty
\label{Le5d}
\end{equation}
and the energy~(\ref{Le5}) is positive.
    For space like vector~$\zeta$, as it take place in the ergosphere of
rotating black hole, the arbitrary positive and negative values are
possible (see~\cite{LPPT}, p.~325).

    Note that in spite of the invariance of the scalar product~(\ref{Le5})
the value~(\ref{Le4}) of the energy depends on the choice of the
reference frame.
    This occurs due to the fact that by changing the reference frame in
which the physical measurements are made the observer is changing
vector~$\zeta$.
    The analysis of the situation in rotating coordinate system in flat
space-time is made in~\cite{GribPavlov2016d}.

    In Minkowski space-time in Galilean coordinate system or any other
coordinate system with $g_{00}=1$, $g_{0\alpha}=0$, ($\alpha =1,2,3$)
the energy~(\ref{Le3}) is
    \begin{equation}
E= c^2 \frac{d t}{d \lambda}
\label{Lee6}
\end{equation}
    and it is always positive in movement ``forward'' in time
because in the future light cone one has $d t/d \lambda > 0$.

    Consider the coordinate system in which metric of flat space-time has
the form of the metric of the expanding homogeneous isotropic Universe
--- Milne universe~\cite{Milne}
    \begin{equation}
d s^2 = c^2 d t^2 - c^2 t^2 \left( d \chi^2 + \sinh^2 \chi d \Omega^2 \right),
\label{Le6}
\end{equation}
    where $d \Omega^2 = d \theta^2 + \sin^2 \theta \, d \varphi^2 $,
coordinate $\chi $ is changing from 0 to $ + \infty $.
    In new coordinates
    \begin{equation}
T = t \cosh \chi, \ \ \ \ r = c t \sinh \chi , \ \ \ \ cT > r > 0
\label{Le7}
\end{equation}
    the interval~(\ref{Le6}) takes the form of Minkowski interval
    \begin{equation}
ds^2 = c^2 d T^2 - d r^2 - r^2\, d \Omega^2.
\label{Le8}
\end{equation}
    This space-time with coordinate $ t\ge 0$, $\chi \ge 0$
corresponds to the future cone in coordinates $cT, r$.

    The radial distance between points $\chi = 0$ and $\chi$ in
metric~(\ref{Le6}) is $D = c t \chi$.
    Taking~$D$ as the radial coordinate~\cite{Ellis93} one obtains
the interval as
    \begin{equation}
ds^2 = \left( 1- \frac{D^2}{c^2 t^2} \right) c^2 d t^2 + 2 \frac{D}{t} dt d D -
d D^2 - c^2 t^2 \sinh^2 \! \left( \frac{D}{ct} \right) d \Omega^2.
\label{Le9}
\end{equation}
    From the condition $ds^2 \ge 0$ one obtains that if $D$ is larger than
$D_s = ct $, then no physical object can be at rest in
coordinates $t, D, \theta, \phi$.
    The value $D_s$ corresponds to $\chi = 1$ and it plays the role of the
static limit for the rotating black hole in
Boyer-Lindquist~\cite{BoyerLindquist67} coordinates.

    The energy of the particle with mass~$m$ in
coordinates $t, D, \theta, \phi$ is
    \begin{equation}
E = c g_{0k} \frac{d x^k}{d \lambda}
= m c^2 \frac{d t}{ d \tau} \left(
1 - \frac{D^2}{c^2 t^2} + \frac{D}{c^2 t} \frac{d D}{d t} \right)
= m c^2 \frac{d t}{ d \tau} \left( 1 + \chi t \frac{ d \chi}{d t} \right).
\label{Le10}
\end{equation}
    From~(\ref{Le6}) one obtains for any physical object the inequality
    \begin{equation}
t \left| \frac{ d \chi}{d t} \right| \le 1 .
\label{Le11}
\end{equation}
    So particle can have negative energy only for $\chi >1$, i.e. out of
the static limit, if
    \begin{equation}
\frac{ d \chi}{d t} < -\frac{1}{\chi t}.
\label{Le12}
\end{equation}

\begin{figure}[ht]
\centering
\includegraphics[width=64mm]{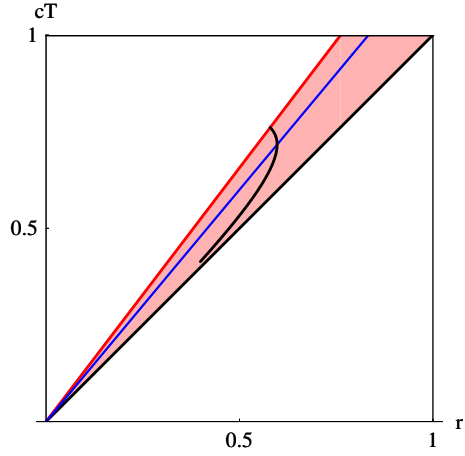}
\caption{Possible region of movement of particle with negative and zero
energies in the reference frame $t, D $ in flat coordinate $T,r$.}
\label{FigM1}
\end{figure}

    Note that the components of metric~(\ref{Le9}) depend on time and
the energy~(\ref{Le10}) in general is not conserved on the geodesics.
    If the energy is zero then particle is moving noninertial according
to the law
    \begin{eqnarray}    \label{Le13}
\frac{ d \chi}{d t} = -\frac{1}{\chi t} \ \ \Leftrightarrow  \ \
\chi = \sqrt{\chi_0^2 - 2 \log (t/t_0 ) }, \ \
t \in \left[ t_0, \ t_0 \exp ((\chi_0^2 -1) /2 ) \right].
\end{eqnarray}
    The trajectory of such movement for case $\chi_0 =2$, $t_0=0.11$ is
represented by the curve on Fig.~\ref{FigM1} in coordinates $T$, $r$
(see~(\ref{Le7})).
    In case of the inertial movement trajectory in these coordinates is
a straight line.
    Possible region of movement of particles with negative and zero energies
in the reference frame $t,D$ is defined in the coordinate $T,r$ by
conditions $1 \le cT/r \le \coth 1 \approx 1.313$.

    Velocities of movement in coordinates $T,r$ and $t, \chi$ satisfy
condition~\cite{GribPavlov2021}
    \begin{equation}
t \frac{d \chi}{d t} = \frac{ \displaystyle
\frac{dr}{d T} - c \tanh \chi}{ \displaystyle c - \frac{d r }{d T} \tanh \chi}.
\label{Le14}
\end{equation}
    So for
    \begin{equation}
\chi \tanh \chi \ge 1
\label{Le15}
\end{equation}
    particles at rest in inertial frame $T,r$ will have negative energies
in the frame $t, D$.
    This region can be seen on Fig.~\ref{FigM1} as the region above the blue
line in red district.
    Zero energy of the particle at rest in $T,r$ coordinates is possible
only the blue line defined by the root of equation
$\chi \tanh \chi = 1$, i.e. $\chi \approx 1.1997$.

    So one can see that for specific choice of coordinates one can obtain
negative and zero energies for particles at rest in inertial frame.


    Note that for small distances ($D/(ct) =\chi \ll 1$) the metric~(\ref{Le9})
becomes the metric of comoving spherical coordinate system of Minkowski
space-time
    \begin{equation}
ds^2 = c^2 d t^2 - d D^2  -D^2 d \Omega^2,
\label{Le9mD}
\end{equation}
    and the energy~(\ref{Le10}) will be equal to usual energy
in inertial system of flat space-time
    \begin{equation}
E_u = m c^2 \frac{d t}{ d \tau} \approx m c^2 \frac{d T}{ d \tau},
\label{Le10mD}
\end{equation}
    because for $\chi \ll 1$ one has $t \approx T$.

    The decay of the body on two bodies, one with negative energy and
the other with the positive energy being larger than the energy of the initial
decaying one corresponds to the Penrose process.
    This process occurs out of the static limit on distance $D > ct$.
    However later these two products of the decay move inside the static limit
and during flight in the direction of the origin where the metric is that
of Minkowski space change their energies in such manner that the result will be
the same as in inertial frame.
    Really due to~(\ref{Le10})
    \begin{eqnarray}
E = E_u + m c^2 \chi t \frac{ d \chi}{d \tau} .
\label{Le10r}
\end{eqnarray}
    Here~$E_u$ is the energy in the reference frame $t,\chi$,
such that $g_{0i} =0$, $i\ne 0$, and $g_{00}$ does not depend on time.
    So $E_u$ is conserved.
    At the point of decay both energies $E$ and  $E_u$ are conserved.
    When the body~2 with the positive energy arrives to the coordinate
origin $\chi=0$ its energy~$E$ (\ref{Le10r}) will be equal to~$E_u$
and no growth of the energy will be observed.

    Body~1 with the negative energy~$E$ due to~(\ref{Le10}) after decay will
have the negative value of velocity $d\chi/ d t$ larger (in absolute value)
than that of the body~2.
    This means that it will arrive to the origin before the arrival of body~2.
    It's energy in the origin of the coordinate frame will be also positive
and the full energy of 1 and 2 will be equal to that of decaying body.
    So at the origin one will not observe any effect which makes this
situation similar to the situation for Kerr's black hole.

    Really for rotating black hole~\cite{Penrose69} as in case of rotating
coordinate system~\cite{GribPavlov2019} the energy is conserved.
    In this case when body~2 will go out of the ergosphere far from it
the body~1 with negative energy moves further inside the horizon of
the black hole or goes to the infinity in case of rotating coordinates
in Minkowski space.
    So the energy of the body with positive energy due to conservation of
the energy will be always larger then that of the initial decaying body.

    Note that existence of states with negative energies for Milne's metric
leads to effect similar to Hawking effect~\cite{Hawking75} for Schwarzschild
metric.
    Particle creation in quantum theory will occur and the detector of
particles checks them (see Sec.~5.3 in~\cite{BirDev}).
    This will be creation of virtual particles (see~Sec.~9.8 in~\cite{GMM})
so no change of the metric due to them can be observed.

\vspace{4mm}
\section{\normalsize Negative energy in G\"{o}del universe}
\label{secGode}

\hspace{\parindent}
    Metric of the G\"{o}del cosmological model of the rotating Universe
proposed in 1949 (see~\cite{Godel49} or~\cite{Godel})
can be written as
    \begin{equation}
d s^2 =  c^2 d t^2 - d x_1^2 +
\frac{\exp \left( 2\sqrt{2} \omega x_1/ c \right) }{2} dx_2^2
+ 2 \exp \left( \sqrt{2} \omega x_1 /c \right) c d t dx_2 - d x_3^2 ,
\label{g1}
\end{equation}
    where $\omega$ is constant.
    Such metric is the exact solution of Einstein's equation with background
matter as ideal liquid without pressure and negative cosmological
constant~$\Lambda $
     \begin{equation}
R_{ik} -\frac{1}{2} R g_{ik} + \Lambda g_{ik} = - 8 \pi \frac{G}{c^4} T_{ik},
\label{Eur}
\end{equation}
    were
    \begin{equation}
- \Lambda = \left( \frac{\omega}{c} \right)^2 = 4 \pi \frac{G}{c^2} \rho,
\ \ \ \ T_{ik}= \rho c^2 u_i u_k,
\label{gT1}
\end{equation}
$u^i = \delta^i_0$.
     Here $\omega$ has the sense of the angular velocity of rotation of
the vector of fluid of the background matter~$u^i$.

    Taking instead of $t,x_1,x_2$ new coordinates $t',r,\phi$:
    \begin{equation}
\exp \left( \sqrt{2} \omega x_1 /c \right) = \cosh 2 r + \cos \phi \sinh 2 r,
\label{g2}
\end{equation}
    \begin{equation}
\omega x_2 \exp \left( \sqrt{2} \omega x_1 /c \right) = \sin \phi \sinh 2 r,
\label{g3}
\end{equation}
    \begin{equation}
\tan \frac{1}{2} \left( \phi + \omega t - \sqrt{2} t' \right)=
\exp(-2 r) \tan \frac{\phi}{2} ,
\label{g4}
\end{equation}
    one writes the interval~(\ref{g1}) in the form~\cite{Godel49,HawkingEllis}:
    \begin{equation}
ds^2 = 2 \left( \frac{c}{\omega} \right)^{2} \left( dt'^2 - d r^2 +(\sinh^4 r
- \sinh^2 r) d \phi^2 + 2 \sqrt{2} \sinh^2 r d \phi d t' \right) - dx_3^2,
\label{g5}
\end{equation}
    where
$-\infty < t' < \infty$, $0\le r < \infty$, $0 \le \phi < 2 \pi $
with identifying $\phi =0$ and $\phi = 2 \pi$.

    Now consider the general case of space-time $t',r,\phi,z$ with the interval
    \begin{equation}
d s^2 = a^2\! \left[ \left( d t' + \Phi(r) d \phi \right)^2\!\! - d r^2\! - d z^2 \!
- R^2(r) d \phi^2 \right],
\label{ge1}
\end{equation}
    where $a$ is constant,
$-\infty < t < \infty$, $0\le r < \infty$, $-\infty < z < \infty$,
$0 \le \phi \le 2 \pi $ with identifying $\phi =0$ and $\phi = 2 \pi$.
    Let's $ \Phi(r) >0$ and $R(r) > 0$ for $r>0$.
    For G\"{o}del universe $a= \sqrt{2} c /\omega$, $z= x_3/a $ and
    \begin{equation}
\Phi(r) = \sqrt{2} \sinh^2 r , \ \ \ \
R(r)= \sinh r \cosh r.
\label{ge2}
\end{equation}
    The metrical tensor is
    \begin{equation}
(g_{ik} ) = a^2 \left(
\begin{array}{cccc}
1 & \Phi & 0 & 0 \\[4pt]
\Phi \ \ & \Phi^2 - R^2 & 0 & 0 \\[4pt]
0& 0 & -1 & 0 \\[4pt]
0& 0& 0& -1
\end{array} \right),
\label{geik}
\end{equation}
    \begin{equation}
(g^{ik} ) = \frac{1}{a^2 R^2(r)} \left(
\begin{array}{cccc}
R^2 - \Phi^2 & \Phi & 0 & 0 \\[7pt]
\Phi & - 1 & 0 & 0 \\[4pt]
0& 0 & -R^2 & 0 \\[4pt]
0& 0& 0& -R^2
\end{array}
\right),
\label{geikO}
\end{equation}
    where indexes $i,k = 0,1,2,3$ correspond to $t',\phi,r,z$.
    Note that for any $r>0$ the metrical tensor is not degenerate
$ {\rm det}\,(g_{ik}) = -a^8 R^2(r) < 0 $.
    The degeneration for $r=0$ in G\"{o}del universe is coordinate degeneracy.
    The eigenvalues of $g_{ik}$ tensor are
    \begin{eqnarray}
\lambda_{1,2} &=& \frac{a^2}{2} \left( \Phi^2 - R^2 + 1
\pm \sqrt{ (\Phi^2 - R^2 +1)^2 + 4 R^2} \right), \nonumber \\
\lambda_{3,4} &=& - a^2.
\label{gsz}
\end{eqnarray}
    For $r>0$ one has
    \begin{equation}
\lambda_{1} \ge a^2, \ \ \ \
0 > \lambda_2 \ge - a^2 R^2.
\label{gsz12}
\end{equation}
    Note that inspite $g_{\phi \phi}$ is positive for $\Phi(r) > R(r)$
the signature of $g_{ik}$ for all $r>0$ is the standard $(+,-,-,-)$.

    Possible movement of particles is defined by $ds^2\ge0$
so for the interval~(\ref{ge1}) one has
    \begin{equation}
d t'^2 + \left( \Phi^2(r)- R^2(r)\right) d \phi^2 + 2 \Phi(r) d \phi d t'
- d r^2 - d z^2 \ge 0 .
\label{ge1n}
\end{equation}
    It is important that for any coordinate system with interval~(\ref{ge1})
the physical body for any values of $ r, \phi, z$ can be at rest, i.e.
there is no static limit!
    However in~(\ref{ge1}) there is nondiagonal term $d t' d \phi$ like in
Kerr's metric.
    But differently from the case of rotating coordinate
system~\cite{GribPavlov2016d} there is the possibility of the change of
the sign before $d\phi^2$.

\begin{figure*}[ht]
\centering
\includegraphics[width=50mm]{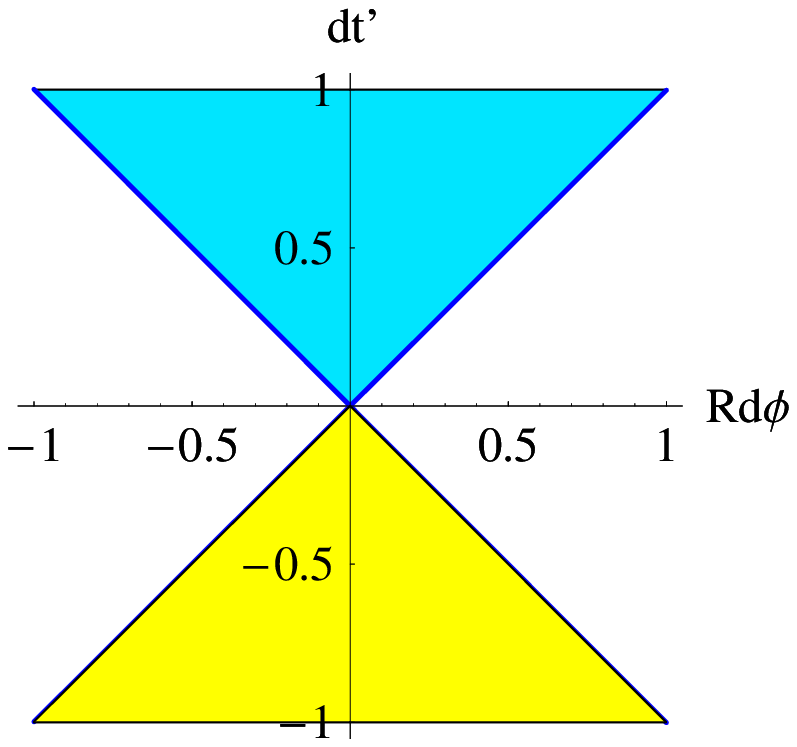} \
\includegraphics[width=50mm]{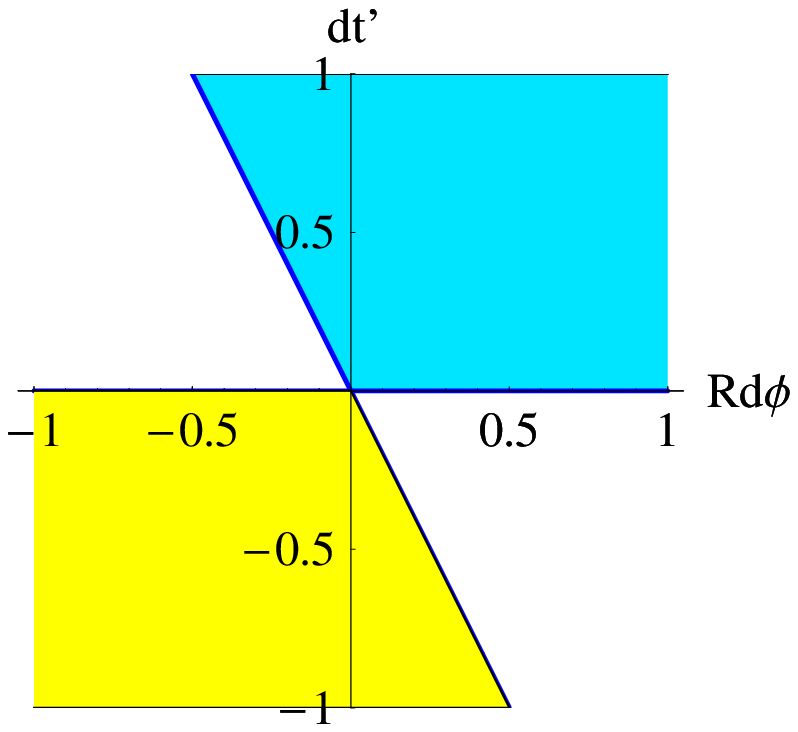} \
\includegraphics[width=50mm]{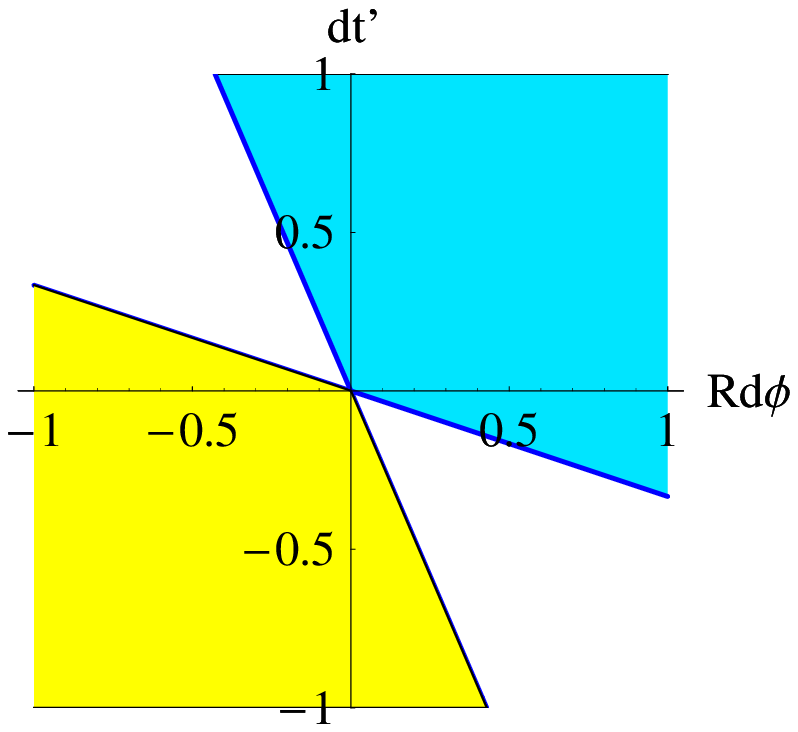}
\caption{Light ``cones'' of future (blue color) and past (yellow color) for
G\"{o}del universe in coordinates $t',\phi$ for cases $r=10^{-3}$ (left),
$r=r_0$ (center) and $r=2 r_0$ (right).}
\label{Fig1}
\end{figure*}

    From~(\ref{ge1n}) one obtains
    \begin{equation}
\left( \frac{d t'}{d \phi} \right)^2 + 2 \Phi(r) \frac{d t'}{d \phi}
+ \Phi^2(r)- R^2(r) \ge 0 .
\label{gee1}
\end{equation}
    The solution of this inequality is the union of two intervals
    \begin{equation}
\frac{d t'}{d \phi} \in \left( -\infty, -(\Phi(r) + R(r)) \right] \cup
\left[ R(r) - \Phi(r), +\infty \right).
\label{gee2}
\end{equation}
    Considering cases of different signs of $d\phi$, one obtains
the following sets of solutions of~(\ref{gee1}):
    \begin{equation}
    d \phi \ge 0 \Rightarrow
\begin{array}{l}
dt' \ge (R-\Phi) d \phi \ \vee \ d t' \le -(R+\Phi) d \phi,
\end{array}
\label{gee3}
\end{equation}
    \begin{equation}
    d \phi \le 0  \Rightarrow
\begin{array}{l}
dt' \ge - (R + \Phi) d \phi \ \vee \ d t' \le (R - \Phi) d \phi.
\end{array}
\label{gee4}
\end{equation}
    These sets define light ``cones'' of future and past for
the metric~(\ref{ge1}).
    The form of these cones in cases $\Phi \ll R$, $\Phi =R$
and $\Phi > R$ is shown in Fig.~\ref{Fig1} for the G\"{o}del universe with
    \begin{equation}
\Phi(r) > R(r) \ \ \Leftrightarrow \ \ r > r_0=\log(1 + \sqrt{2} ).
\label{gn5}
\end{equation}

    Let us find limitations on possible values of the energy of particles
moving in such universe.
    The coordinate~$t'$ is dimensionless, so the ``physical energy'' of
the particle is expressed through the time component of the momentum as
    \begin{equation}
E = p_0 \frac{c}{a} = g_{0k} \frac{c}{a} \frac{d x^k }{d \lambda}.
\label{gi0}
\end{equation}
    For the frame with coordinates~(\ref{ge1}) the covariant $t',\phi,z$
components are conserved, because the component of metric depend only on~$r$.
    So the conserved energy on the geodesic for the interval~(\ref{ge1}) is
    \begin{equation}
E = c a \left( \frac{ d t'}{d \lambda} + \Phi(r) \frac{ d \phi}{d \lambda}
\right) .
\label{gi0k}
\end{equation}

    From~(\ref{gee3}), (\ref{gee4}) for the case of movement ``forward''
in time, i.e. in the future light cone one obtains
    \begin{equation}
dt' + \Phi d \phi \ge R |d \phi |,
\label{gee5}
\end{equation}
    so
    \begin{equation}
E \ge c a  R \frac{|d \phi |}{d \lambda}.
\label{gee6}
\end{equation}
    It means that for particle moving in the future cone in G\"{o}del universe
the energy is not negative.

    For movement ``back in time'' the energy is limited from above by
    \begin{equation}
E \le - c a  R \frac{|d \phi |}{d \lambda}
\label{gee7}
\end{equation}
    and so it can be less or equal zero.
    However such movement physically is inconsistent.
    The ``time machine'' effect in G\"{o}del universe corresponds to continuous
movement in the future cone.
    So for $r > r_0$, where $\Phi(r) > R(r)$ closed loops
(they are not geodesic lines) $r={\rm const}$, $ z = {\rm const}$,
called G\"{o}del cycles, from $\phi= 0$ to $ \phi = 2 \pi$
are closed time-like curves~\cite{HawkingEllis}.
    Particle moving along such cycle is moving ``forward'' in time but due to
identification of values $\phi$ different on $2 \pi$
it occurs in the past after the whole cycle.
    It's energy is positive due to~(\ref{gee6}).
    Such ``time machine'' is different from that moving in the past by the
sign of particle energy.

\vspace{4mm}
\section{\normalsize Conclusion}
\label{secConcl}

   Three different cases are investigated concerning the possibility
of existence of particles with negative and zero energies.

    1. Schwarzschild black hole of the mass~$M$.
    Trajectories of particles with negative and zero energies exist inside
of the horizon of black hole which can be shown in Kruskal--Szekeres
coordinates.

    2. Flat space-time in Milne's coordinates.
    Here one also has the possibility of existence of particles with
negative and zero energies if nonsynchronous system of coordinates
is used.

    3. G\"{o}del cosmological model with rotation.
    Here we proved that in this model in G\"{o}del's coordinates
particles with negative and zero energies don't exist.

    As to observations of the discussed effects one can say that even so
well established Penrose and Hawking effects are still not observed.
    However we hope that in future development of  observational astrophysics
one can see the  consequences of existence of negative and zero energies.

\vspace{7pt}
\noindent
{\bf Author Contributions:} A.A.G. and Y.V.P. have contributed equally to
all parts of this work. All authors read and agreed
to the published version of the manuscript.

\vspace{7pt}
\noindent
{\bf Funding:}
This research is supported by the Russian Science Foundation (Grant No. 22-22-00112).

\vspace{7pt}
\noindent
{\bf Conflicts of Interest:} The author declares no conflict of interest.


\end{document}